\documentclass[useAMS,usenatbib]{mn2e}
\usepackage{graphicx, epsfig}
\title[Void alignment and density profile applied to measuring cosmological parameters]{Void alignment and density profile applied to measuring cosmological parameters}
\author[De-Chang Dai]{De-Chang Dai$^{1}$\thanks{E-mail:
diedachung@gmail.com}
\\
 Institute of Natural Sciences, Shanghai Key Lab for Particle Physics and Cosmology, \\
and Center for Astrophysics and Astronomy, Department of Physics and Astronomy,\\
Shanghai Jiao Tong University, Shanghai 200240, China}

\begin{document}

\date{}

\pagerange{\pageref{firstpage}--\pageref{lastpage}} \pubyear{}

\maketitle

\label{firstpage}

\begin{abstract}
{ We study the orientation and density profiles of the cosmological voids with SDSS10 data. Using voids to test Alcock-Paczynski effect has been proposed and tested in both simulations and actual SDSS data. Previous observations imply that there exist an empirical stretching factor which plays an important role in the voids' orientation. Simulations indicate that this empirical stretching factor is caused by the void galaxies' peculiar velocities. Recently Hamaus et al. found that voids' density profiles are universal and their average velocities satisfy linear theory very well. In this article we first confirm that the stretching effect exists using independent analysis. We then apply the universal density profile to measure the cosmological parameters. We find that the void density profile can be a tool to measure the cosmological parameters. }
\end{abstract}


\section{introduction}
Observations have shown that the universe is isotropic and homogeneous at large scales. At the same time, the universe is highly anisotropic at smaller scales and builds up a hierarchical structure of matter from galaxies to super clusters. While these dense structures are forming, it is unavoidable to create underdense regions which are called voids. Like the overdense regions, the voids are also highly affected by the evolution of the universe's energy density components and therefore could be a powerful cosmological probe\citep{Thompson2011}. For example, their shapes and sizes are sensitive to the nature of the dark energy \citep{Biswas,Bos2012,Jennings2013,Li2012} and their internal dynamics may represent the fifth force in the modified gravity theory\citep{2015MNRAS.451.1036C,Clampitt2013,Li2009}.

Unlike luminous overdense regions, the voids can not be found directly, partly because they are generally faint (because of the lack of galaxies) or even completely dark. Thus, they are identified through mapping of the overdense areas first and then by using void finders\citep{Sutter:2014haa} to pick up the voids.  At large scales, the voids' shapes are statically isotropic and therefore stacking could make the voids look like standard spheres. This is the basic requirement to precede the  Alcock-Paczynski test\citep{Alcock}, and several such tests have been done \citep{Sutter2012a,Sutter:2013ssy,Sutter:2014oca} either through the Millenium Run simulation \citep{2005Natur.435..629S} or actual observations.

Since voids' shapes have been affected by the evolution of the cosmological components, their orientation will also be affected by the evolution. Thus, we could also use orientation of the voids to test a gravitational model in question. Simulations have shown that the orientations of voids are highly correlated with their close neighbors (within $30 \mbox{Mpc}$)\citep{Platen:2007ng,Lee:1999ii,Porciani:2001er,Lee:2006gj,Park:2007qc}. This, however, has not been confirmed in any actual observation. In this article we are not trying to test this short distance alignment, instead we are focusing on the large scale orientation. At such scales Alcock-Paczynski effect and galaxies peculiar velocities are the main source of distortion. Our study confirms that empirical stretching factor (which is caused by peculiar velocities) also plays an important role in the orientation of the voids and their density profiles. As a consequence, by studying the void density profile, we could study the cosmological energy density components. In the following section we first describe our dataset from the public Cosmic Void Catalogue \citep{Sutter2012,Sutter:2013pna,Sutter:2014oca} and then show the distribution of the orientation of the void. The Alcock-Paczynski effect strongly affects the statistics of the long axis and could be a good probe to test the cosmical parameters.  The orientation test is similar to the one by Sutter et al. \citep{Sutter:2014oca}, though the method is different. However, this is not enough to quantify the peculiar velocity effect. Therefore, we fit the void density with an empirical universal density. This in turn could quantify the peculiar velocity effect.  From the $\chi^2$ and likelihood analysis change with $\Omega_\Lambda$, we find that the density profile can serve as the cosmological probe, though the current void number might not be fully sufficient yet for very precise results and a prior function must be known to study the confidence level.

\section{Dataset}
Recently, several void catalogues were completed and released\citep{Nadathur:2013bba,Sutter2012}. In this study, we use the voids in the Public Cosmic Void catalogue\citep{Sutter2012,Sutter:2013pna,Sutter:2014oca}, because this catalogue includes the most recently results. This Catalogue puts the voids and void galaxies in the following coordinates

\begin{eqnarray}
x_1&=&D_c(z)\cos(\textrm{Dec})\cos(\textrm{Ra})\nonumber\\
\label{redshift-coordinate}
x_2&=&D_c(z)\cos(\textrm{Dec})\sin(\textrm{Ra})\\
x_3&=&D_c(z)\sin(\textrm{Dec})\nonumber
\end{eqnarray}

Here $z$ is redshift. $D_c(z)$ is the comoving distance from the center. Ra is Right ascension and Dec is Declination in the equatorial coordinate system.  Since the data set is within $z<1$, the radial component can be ignored, and $D_c(z)$ is written as

\begin{eqnarray}
D_c&=&\frac{c}{H_0}\int_0^z\frac{dz'}{E(z')}\\
E(z)&=&\sqrt{\Omega_m(1+z)^3+\Omega_\Lambda}
\end{eqnarray}

Here, we assume the universe is flat. $\Omega_\Lambda$ is the dark energy component and $\Omega_m$ is the matter component. $\Omega_\Lambda+\Omega_m=1$. The catalogue includes two sub-catalogues (comoving and redshift) according to the coordinate in which it is searched for the voids.  Voids in comoving catalogue are found out through assuming dark matter component $\Omega_\Lambda =0.73$ and Voids in redshift catalogue is found out through assuming dark matter component $\Omega_\Lambda =1.$. The void number and redshift range in each section are listed in table \ref{tab:samples}. We use only the "central" voids which are selected to avoid the boundary or mask effect\citep{Sutter2012a,Sutter:2013pna,Sutter:2014oca}.

\begin{table}
\centering
\caption{The void number in different section}
\tabcolsep=0.11cm
\footnotesize
\begin{tabular}{cccccc}
  Sample Name &  $z_{\rm min}$ &
              $z_{\rm max}$  &
              redshift & comoving\\
   &   &  & $N_{\rm void}$ & $N_{\rm void}$\\
  \hline  \hline
dr72dim1  & 0.0 & 0.05  &  110 &102\\
dr72dim2  & 0.05 & 0.1  & 186 &184\\
dr72bright1 & 0.1 & 0.15  & 186 &188\\
dr72bright2 & 0.15 & 0.2  & 112 &96\\
dr10lowz2  & 0.2 & 0.3  &  138&136\\
dr10lowz3 & 0.3 & 0.4  &  215&198\\
dr10lowz4  & 0.4 & 0.45  & 71 &90\\
dr10cmass1 & 0.45 & 0.5  &  227&229\\
dr10cmass2  & 0.5 & 0.6  & 694 &696\\
dr10cmass3  & 0.6 & 0.7  &  386&$\times$\\
\hline
\end{tabular}
\label{tab:samples}
\end{table}

\section{Testing isotropy at large scales}
The void alignment depends on three sources: initial conditions, void packing and tidal forces\citep{Platen:2007ng}. These three sources  should be local effects and should not contribute much at large scales. Therefore the distribution of the orientation of the cosmological voids should be a random distribution. Based on this, it was suggested that the void orientation with respect to the line of sight could be a measure of the redshift distortion\citep{Foster2009,Ryden:1995tc}.

If void orientation is random, the angle, $\theta$, between its orientation direction and line-of-sight satisfies the distribution

\begin{equation}
P(\theta)d\theta=\sin\theta d\theta
\end{equation}

or
\begin{equation}
P(\cos\theta)d\cos\theta=d\cos\theta
\end{equation}

To examine this assumption, we calculate the orientation of voids through ellipsoid approximation. We compute the location of the  void's center through the volume weight as

\begin{equation}
\vec{x}_c=\frac{\sum \vec{x}^i V_i}{\sum V_i}
\end{equation}

Here, $\vec{x}^i$ is the position of the void's i-th galaxy's and $V_i$ is its Voronoi cell's volume. We then compute the shape tensor, $S_{ij}$,

\begin{equation}
S_{ij}=\sum_k (x_i^k-{x_c}_i)(x^k_j-{x_c}_j)m_k
\end{equation}

\begin{figure}
   \centering
\includegraphics[width=9cm]{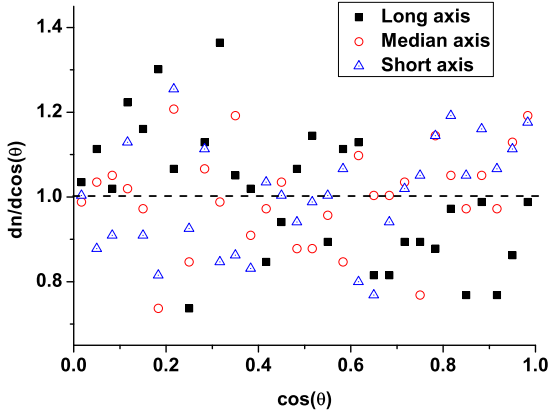}
\caption{The three axes' orientation distribution. We set $\Omega_\Lambda=0.73$ for the figure.  We use voids in the comoving catalogue in the 9 sections. The orientation of the long axes are slightly anti-aligned to the line-of-sight. The bin size of $\cos\theta$ is 1/30.}
\label{iso-full}
\end{figure}

$k$ is the void galaxy's index. $i$ and $j$ are the three position's indexes. $m_k$ is the mass weight. Here we choose $m_k=1$ for all the cases. The three eigenvectors are the orientation of the ellipsoid's axes. The three eigenvalues are their axes' length square. We calculate the orientations under $\Omega_\Lambda = 0.73$ to compare with recent WMAP and Planck's measurement\citep{komatsu,Planck2013}.  We use voids in the comoving catalogue and show the result in figure \ref{iso-full}. At first glimpse, one finds that the long axes have values larger than 1 for small $\cos\theta$. That means the long axes are anti-aligned to the line-of-sight. It is not very clear whether median and short axes have any preferred direction without a detailed analysis.

There are at least two reasons to this anti-alignment phenomenon. One is Alcock-Paczynski effect\citep{Alcock} and the other is peculiar velocity. We will discuss them in the next section.

\section{Testing Alcock-Paczynski effect on voids' long axis and peculiar velocity as a cause of stretching}

It is known that Alcock-Paczynski effect could cause redshift distortion. At the same time it also distorts the orientation of a void.
For our purpose, to study the Alcock-Paczynski effect on the orientation, we can't approximate the void shape with a sphere, because of the voids' irregular shapes and their tracers' peculiar velocities. To simplify the argument, we consider the void to be a one dimensional rod. The rods' directions are the orientation of long axes. The real situation is more complicated, but our approximation can at least tell us what is going on when considering the Alcock-Paczyski effect on the void's orientation.

Let's consider a one dimensional void. Its angular extent is $\delta \theta_v$ and redshift extent is $\delta z_v$. In the comoving coordinate system,

\begin{eqnarray}
\delta r_c &=&\delta \theta_v D_c\\
\delta d_c &=&\delta z_v \partial_z D_c
\end{eqnarray}

$\delta d_c$ void's comoving size in line-of-sight direction and $\delta r_c$ is the void comoving size in angular direction. Therefore, the angle between the void's long axis and line-of-sight is

\begin{equation}
\tan\theta_c = \frac{D_c}{\partial_z D_c}\frac{\delta \theta_v}{\delta z_v}
\end{equation}

 $\theta_c$ follows a random distribution \ if $D_c$ is the real comoving distance. If one uses the other factor, $D_n$, ($D_n=\frac{cz}{H_0}$ in general discussion), one will have mistaken the angle between the void's orientation and line-of-sight as

\begin{equation}
\tan\theta_n = \frac{D_n}{\partial_z D_n}\frac{\delta \theta_v}{\delta z_v}
\end{equation}

In the measurement, $\delta z_v$ and $\delta \theta_v$ are measured. However, the comoving distance, $D_c$, is not known directly. One may use $D_n$ instead of $D_c$ to study the voids' orientation. The relation between $\theta_n$ and $\theta_c$ is

\begin{eqnarray}
\label{angle1}
\tan\theta_n = f(z)\tan\theta_c\\
f(z)=\frac{\partial_z \ln D_c }{\partial_z \ln D_n}
\end{eqnarray}

The distribution
\begin{equation}
P(\theta_n)=P(\theta_c)\frac{d\theta_c}{d\theta_n}
\end{equation}

In the actual isotropic coordinate, the void's long axis's are random distributions and therefore $P(\theta_c) = \sin\theta_c$. From equation \ref{angle1}, one finds

\begin{equation}
P(\theta_n)=\frac{\sin\theta_n}{f^2}\Big(\cos^2\theta_n+\frac{\sin^2\theta_n}{f^2}\Big)^{-\frac{3}{2}}
\end{equation}

\begin{figure}
   \centering
\includegraphics[width=8cm]{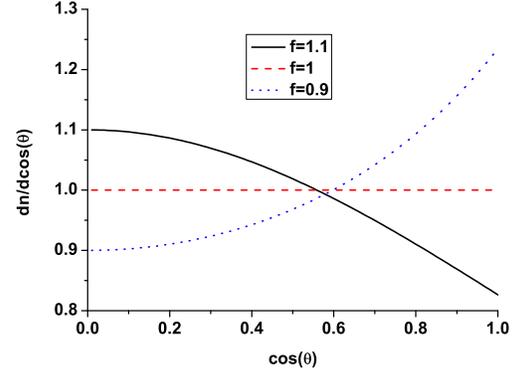}
\caption{$\cos(\theta_n)$ distribution.  It aligns to the line of sight as $f>1$.  }
\label{function-f}
\end{figure}

This is the angular distribution function for a coordinate different from the comoving coordinate system. From figure \ref{function-f}, one finds it is anti-aligned to the line of sight as $f>1$ and aligned to the line of sight  as $f<1$.

\begin{figure}
   \centering
\includegraphics[width=8cm]{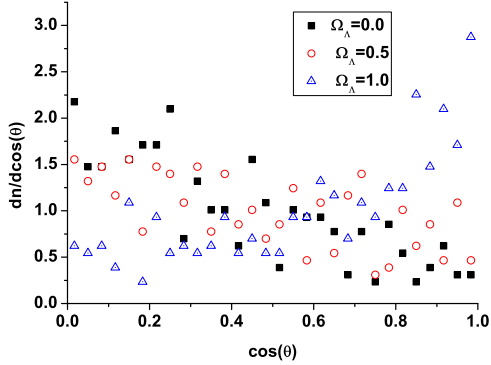}
\caption{The probability function changes with different $\Omega_\Lambda$. Here we use dr10cmass3 dataset.}
\label{long-lambda}
\end{figure}

\begin{figure}
   \centering
\includegraphics[width=9cm]{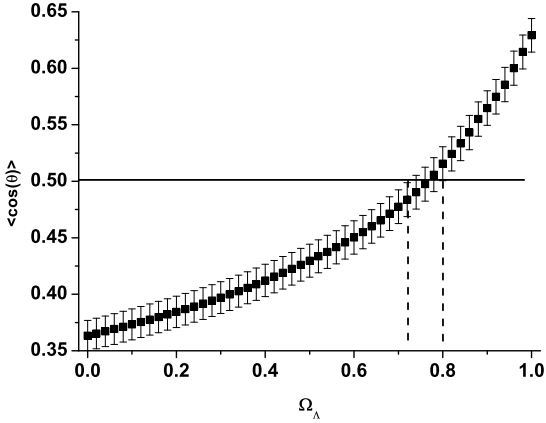}
\caption{$<\cos\theta>$ in different  $\Omega_\Lambda$. The black line is the expected $<\cos\theta>$ in the isotropic case. The vertical dash lines are the $1\sigma$ interval of possible $\Omega_\Lambda$. The data point are redshift types in dr10cmass3 section. }
\label{lambda-change}
\end{figure}

Although the voids' shapes are complicated, we could still  find that the voids' long axis share the same effect. To examine the effect, we use dr10cmass3 dataset because of its higher redshift, and vary $\Omega_\Lambda$ from $0.0$ to $1.0$. Figure \ref{long-lambda} shows that the voids' orientation changes from anti-aligning to aligning to the line-of-sight as we go from $\Omega_\Lambda=0$ to $\Omega_\Lambda=1$ . Based on this, one expects that one could find the dark energy and dark matter component ($\Omega_\Lambda$ and $\Omega_m$) by searching for  $P(\theta_n)$ which best satisfies a random distribution.  For a pure random distribution $P(\theta)$, $<\cos\theta>=0.5$. Therefore, one could calculate the average $<\cos\theta>$ to check whether the distribution follows a random distribution. Figure \ref{lambda-change} shows how $<\cos\theta>$ changes with respect to $\Omega_\Lambda$. The figure shows that $\Omega_\Lambda=0.78\pm 0.02$ in $1\sigma$ interval. However, we must point out that an empirical stretching factor\citep{Sutter:2014oca,Lavaux}, which may be caused by the peculiar velocity, must be included. The same effect has been noticed in the SDSS DR5\citep{Foster2009}. They found that the average orientation angle increases to higher $\theta$ values as $z$ from $0.1$ to $0.16$. If the empirical stretching factor does exist, then $1/f=1.16\pm 0.04$ in a comoving coordinate\citep{Sutter:2014oca,Lavaux} and it will not be isotropic even in the correct comoving coordinate. We can test the empirical stretch factor by calculating the average $<\cos\theta>$.

\begin{equation}
<\cos\theta>=\int_0^{\pi/2}P(\theta)\cos\theta d\theta
\end{equation}

\begin{figure}
   \centering
\includegraphics[width=9cm]{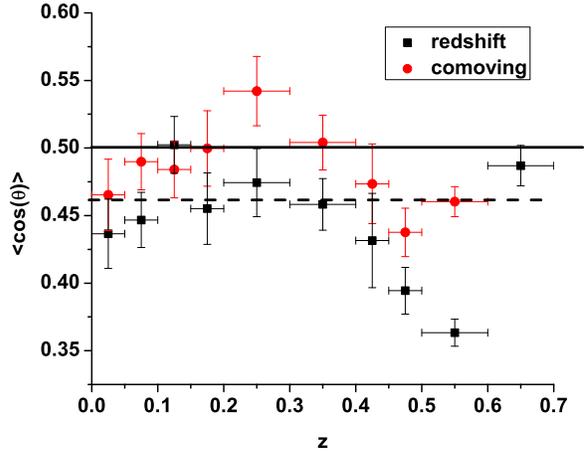}
\caption{$<\cos\theta>$ in different catalogue section. The black one is from the redshift type of the catalogue, and the red one is from the comoving type of the catalogue. $\Omega_\Lambda=0.73$ in the calculation. The black line represents the random distribution result and the dash line represents a stretch factor $1/f=1.16$. }
\label{shift}
\end{figure}

For $1/f=1.16$, $<\cos\theta>=0.46$. We plot all $<\cos\theta>$ in $\Omega_\Lambda=0.73$ case. Figures \ref{shift} shows that most of $<\cos\theta>$ are less than $0.5$, which is the value of a pure random distribution. This confirms the existence of the extra stretching factor. This extra stretch is from the void galaxies' peculiar velocities\citep{Sutter:2014oca,Lavaux}. However, the redshift type of a void and comoving type of a void do not give exactly the same result. This confirms that the void galaxies' quantities  highly depend on the coordinate which is used in the void finder\citep{Nadathur:2013bba}.

\section{Voids' universal density and redshift distortion}

The previous section has confirmed that the voids' orientation depends not only on redshift effect, but also on an empirical factor. According to simulations, the empirical factor is caused by the galaxies's peculiar velocities\citep{Sutter:2014oca,Lavaux}. In general, it is very difficult to study peculiar velocity without any other assumption. However, it has been found that voids' density profile is universal \citep{Hamaus2014,Lavaux,Colberg:2004nd,Padilla:2005ea,Ricciardelli:2013kxa,Ricciardelli:2014vga} and its average velocity fit the linear theory very well\citep{Hamaus2014,Paz:2013sza}. Therefore, we could study the peculiar velocity from this universal density profile.

Here we adopt the density profile form from paper \citep{Hamaus2014}.
\begin{eqnarray}
\frac{\rho_v (r)}{\bar{\rho}}&=&1+\delta_c\frac{1-(r/r_s)^\alpha}{1+(r/r_v)^\beta}\\
v_v&=&-\frac{1}{3}\Omega_m^\gamma H r \Delta (r)\\
\Delta (r)&=&\frac{3}{r^3}\int^r_0 \Big( \frac{\rho_v (q^2)}{\bar{\rho}}-1\Big)q^2 dq
\end{eqnarray}

Here, $\gamma =0.55$. $r=\sqrt{x_1^2+x^2_2+x_3^2}$. $v_v$ is velocity from the linear theory\citep{Peebles}. It has been found that $\alpha$ and $\beta$ at $z=0$ are related to $r_s/r_v$ in the following form\citep{Hamaus2014}.

\begin{eqnarray}
\alpha &=&-2 (r_s/r_v -2)\\
\beta& = &\left\{
  \begin{array}{lr}
    17.5r_s/r_v-6.5 & \mbox{for $r_s/r_v<0.91$}\\
    -9.8r_s/r_v+18.4 & \mbox{for $r_s/r_v>0.91$}
  \end{array}
\right.
\end{eqnarray}

However, we study voids in $z\neq 0$ case and therefore we should  treat them as two free parameters. We hope that $\alpha$ and $\beta$ dependence on the redshift could be found in the future studies. So far,  $x_3$ can not be obtained directly from redshift, and one must consider the distortion effect from the galaxies' peculiar velocities. Therefore,

\begin{eqnarray}
x_3&=&Z-v/H\\
Z&=& \frac{c}{H}\delta z
\end{eqnarray}
where $v$ is the galaxy's velocity and $\delta z$ is the redshift different between galaxy and void center. Since the galaxy's velocity is involved in the measurement,  one needs a velocity marginal function to find $x_3$. The simplest marginal function will be a Gaussian distribution with the center at $v_v$. However, since we rescale the void according to its effect radius, ($R_v= (\frac{3V}{4\pi})^{1/3}$, $V$ is the void's volume) while stacking voids, the velocity will be rescaled too. Therefore whether Gaussian distribution is a velocity distribution function is unclear. We hope future N-body simulation could also provide the form of marginal function. Right now we use Gaussian distribution.

\begin{eqnarray}
f_g(\eta) &=&\frac{1}{\sqrt{2\pi}v_0}\exp(-\frac{\eta^2}{2v_0^2})
\end{eqnarray}

$v_0$ is the root mean square of the velocity. Exponential distribution is also a popular distribution in two-body correlation. We had tested it and found the result is similar to Gaussian distribution one, therefore we are going to give only Gaussian distribution's result. Of course $v_0$ should be a $r$ dependent function. Finding its actual form will rely on the future studies, and here we will treat it as a constant. The average density in $(x_1,x_2,Z)$ is

\begin{eqnarray}
\rho_o(x_1,x_2,Z;\Omega_\Lambda) dZ &=& \int \frac{\rho_v(x_1,x_2,x_3)}{\bar{\rho}} f(w-\frac{x_3}{r}v_v)\nonumber\\
&&\delta(Z-x_3-\frac{x_3}{r}\frac{v_v}{H} ) dx_3 dw
\end{eqnarray}
where $\delta(\eta)$ is a Dirac Delta function.  $\Omega_\Lambda$ appears in the formula, because Hubble constant and position are functions of $ \Omega_\Lambda$. Different $\Omega_\Lambda$s give different mass density. We first rescale the voids given in comoving coordinates according to their effect radius ($R_v$) and then stack them together. We expect a single galaxy's peculiar velocity could be several hundred to thousand $km/s$. This will give several to dozens Mpc of distortion in line-of-sight ($Z$) direction. Therefore we focus only on voids bigger than $10 \mbox{Mpc}$. Low redshift voids either do not satisfy or the void number is very low (less than  one hundred). Considering the void number, we  choose dr10mass2 to study the effect. We use voids in comoving coordinates to minimize the void finder's selection effect.

\begin{figure}
   \centering
\includegraphics[width=9cm]{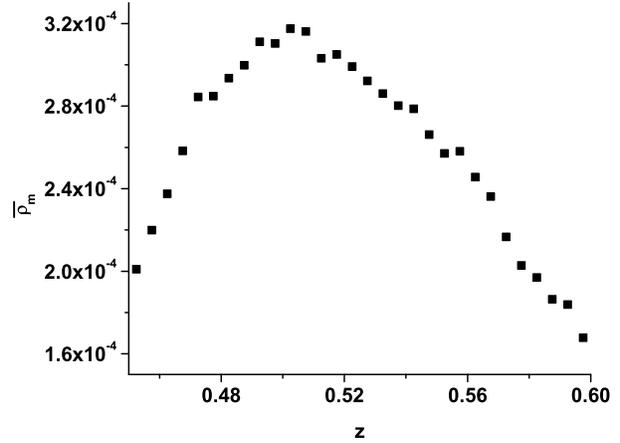}
\caption{The galaxy density according with respect to redshift: The density is calculated according to $\Omega_\Lambda=0.73$. The density depends highly on redshift and can not be treated as a constant. The density unit is $\mbox{Mpc}^{-3}$}
\label{density-z}
\end{figure}

In order to find the density profile, one needs the average galaxy density. This however is not easy, because the average galaxy density depends on survey's capability and is redshift and location dependent.   Figure \ref{density-z} shows how the galaxy density changes with respect to redshift in the comoving space. It is clear the galaxy density highly depends on the redshift. At this moment we treat this density  as average density ($\bar{\rho}_m(z)$) at each redshift\citep{Nadathur:2013bba,Nadathur:2014qja}. We then stack the voids according to void radius, $R_v$.  The density profile is calculated in the following way

\begin{equation}
\bar{\rho}_a(Z,d_v;\Omega_\Lambda) =\frac{1}{N}\sum_i \frac{\rho_i(Z,d_v;\Omega_\Lambda)}{\bar{\rho}_m(z_i)}
\end{equation}

  $\rho_i(Z,d_v;\Omega_\Lambda)$ is the galaxy density of $i$th void at $(Z,d_v)$ respected to $\Omega_\Lambda$. Here $z_i$ is the redshift of the bin. $d_v=\sqrt{x_1^2+x_2^2}$ and $N$ is the number of voids in the stacking. The error of each bin is calculated from

\begin{equation}
\sigma^2(Z,d_v;\Omega_\Lambda)=\frac{1}{N(N-1)}\sum_i (\frac{\rho_i(Z,d_v;\Omega_\Lambda)}{\bar{\rho}_m(z_i)}-\bar{\rho}_a)^2
\end{equation}

\begin{figure}
   \centering
\includegraphics[width=9cm]{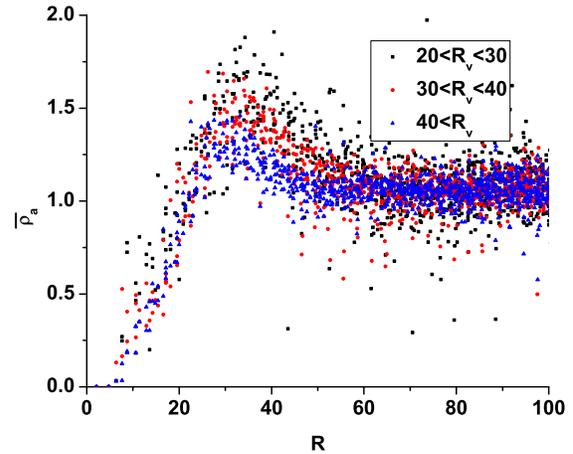}
\caption{The voids are separated into three different radius range ($20\mbox{Mpc}<R_v<30\mbox{Mpc}$, $30\mbox{Mpc}<R_v<40\mbox{Mpc}$ and $40\mbox{Mpc}<R_v$) stacking according to their radius. We rescale all the voids to $30\mbox{Mpc}$ and stack them together. The bin size is $3\mbox{Mpc} \times 3\mbox{Mpc}$ in $(Z,d_v)$ coordinate.}
\label{density-profile}
\end{figure}

 Figure \ref{density-profile} shows the stacking density profile according to different void size.  We stack 175 voids for $R_v>40\mbox{Mpc}$ case, 202 voids for $40\mbox{Mpc}>R_v>30\mbox{Mpc}$ case and 236 voids for $30\mbox{Mpc}>R_v>20\mbox{Mpc}$ case.  The bin size is chosen to be $3\mbox{Mpc}\times 3\mbox{Mpc}$. The stacking void number does not represent the stacking galaxy number in study because the different void has different volume and there are different masked regions to avoid. The smaller voids' profile has a higher density right outside the voids' radius. Apparently the density profile depends on the voids' sizes. Therefore it is unlikely to stack each void without considering its radius. At the following we study void density profile according to these three stacking void profiles. Also we choose $\delta_c=-1$, because there is almost no galaxy near the void center.

\begin{figure}
   \centering
\includegraphics[width=9cm]{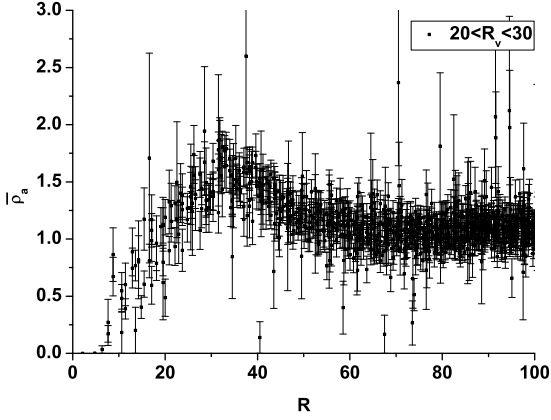}
\caption{ The density distribution with error bar: The stacking voids' effect radius are from $20 \mbox{Mpc}$ to $30 \mbox{Mpc}$. $\Omega_\Lambda=0.5$. Since the limitation of dataset, sometimes the error bar is underestimated. In this case, a point near $R_v=40\mbox{Mpc}$ is much smaller than its neighborhood. This causes that its error bar is underestimated.  }
\label{error}
\end{figure}

  Figure \ref{error} shows average density($\bar{\rho}_a(Z,d_v)$) and error ($\sigma(Z,d_v)$). Near $R_v=40\mbox{Mpc}$, there is a point with much lower $\bar{\rho}_a(Z,d_v)$ and its $\sigma(Z,d_v)$ is much less than its neighborhood. This shows that if the average density ($\bar{\rho}_a(Z,d_v)$) is much smaller than average galaxy density, then $\sigma(Z,d_v)$ is underestimated. If these underestimated points are included in the calculation the increase in $\chi^2$ will be significant. Since these abnormal points appear at small $d_v$ and are caused by the limitation of the dataset, we therefore remove $d_v<3\mbox{Mpc}$ data points to avoid the underestimate of $\sigma(Z,d_v)$.

\begin{figure}
   \centering
\includegraphics[width=9cm]{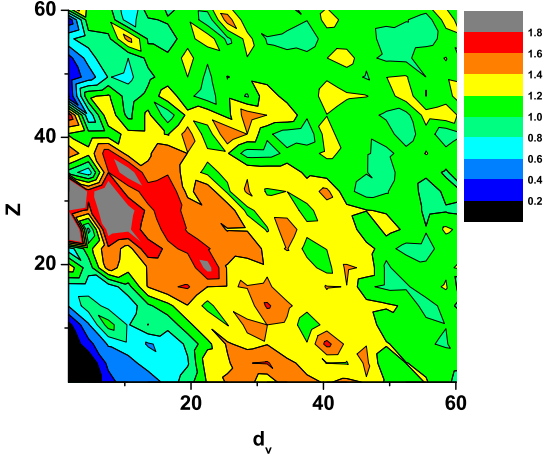}
\caption{ The dr10mass2 density distribution: we stack 400 voids. The voids' effect radius are from $30 \mbox{Mpc}$ to $40 \mbox{Mpc}$. The bin size is $3\times 3 \mbox{Mpc}^3$ in $(Z,d_v)$ coordinate. Here $\Omega_\Lambda =0.73$. We choose $R_e=30\mbox{Mpc}$ in the figure.}
\label{original-map}
\end{figure}

 Figure \ref{original-map} shows the dr10mass2's average density profile in $Z-d_v$ plane. (Here we show $\Omega_\lambda=0.73$) One finds extension in Z-direction at small radius($\sim 10\mbox{Mpc}$) and stretch at larger radius ($\sim 40\mbox{Mpc}$). Highest density peak appears near the Z-axis. This is an evidence of galaxy's peculiar velocity causing the density distribution  distortion.

We then use the universal density profile to fit the stacking data.   We assume each bin is independent from the others. Their cross correlations are independent and diagonal. Therefore we include only $\sigma^2(Z,d_v;\Omega_\Lambda)$ to calculate $\chi^2$.    $\chi^2$ is calculated according to

\begin{eqnarray}
\chi^2(\Omega_\Lambda, \alpha,\beta, v_0, r_s) &=&\sum_{Z,d_v}\frac{(\bar{\rho}_a(Z,d_v)-\rho_o(Z,d_v))^2}{\sigma^2(Z,d_v;\Omega_\Lambda)}
\end{eqnarray}

\begin{figure}
   \centering
\includegraphics[width=9cm]{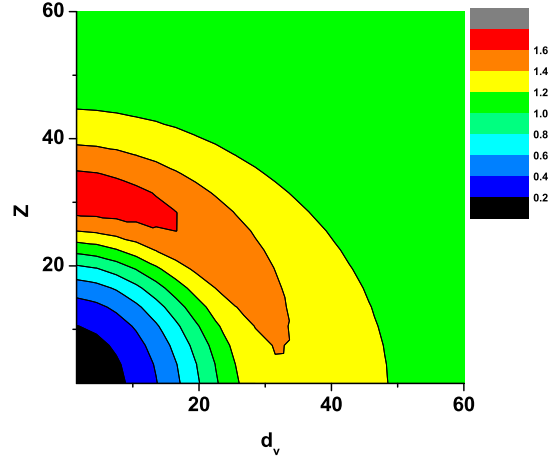}
\caption{The density profile by fitting empirical density function. The other parameters are the same as figure \ref{original-map}.}
\label{fit-map}
\end{figure}

 We use data between $10\mbox{Mpc}$ to $60\mbox{Mpc}$ to calculate $\chi^2$. Figure \ref{fit-map} is the fit of the density profile from the empirical density profile. The result shows similar density stretching, extension and distortion as in figure \ref{original-map}.

\begin{figure}
   \centering
\includegraphics[width=8cm]{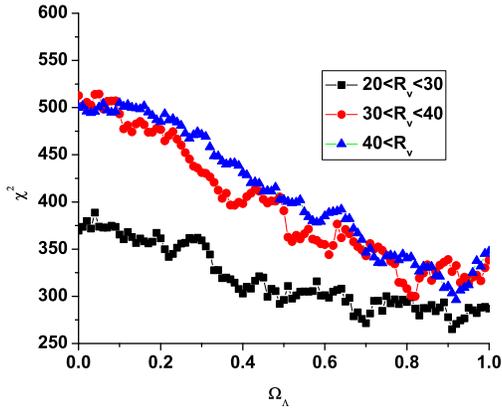}
\caption{ The $\chi^2$ vs. $\Omega_\Lambda$. Here $\alpha$, $\beta$, $v_0$ and $r_s$ are not fixed.}
\label{chi2}
\end{figure}

One of our goals is to find the cosmological dark energy component, $\Omega_\Lambda$.  Figure \ref{chi2} shows the minimum $\chi^2$ changes with respect to $\Omega_\Lambda$ (The other parameters are not fixed). We fit 291 points with 5 parameters. Therefore a reasonable $\chi^2$ value is $\sim 286\pm \sqrt{286}$. $\chi^2$ at small $\Omega_\Lambda$ is farther away from this region and can be ruled out.  In $30\mbox{Mpc}<R_v<40\mbox{Mpc}$ and $40\mbox{Mpc}<R_v$ case, the $\chi^2$ minimums is still a little too large and their best fit parameters are not reasonable ($v_0$ less than $100km/s$ which is unreasonable because the bin size, $3$Mpc, at least induces $300km/s$ uncertainty on velocities). This implies a further study on the form of universal density profile is needed. At the same time, we assume all the points are independent and the likelihood is

\begin{equation}
L(\Omega_\Lambda, \alpha,\beta, v_0, r_s)=\prod\limits_i\frac{1}{(2\pi \sigma_i^2)^{1/2}}\exp(-\chi^2/2)
\end{equation}

 The probability for a particular $\Omega_\Lambda$ is calculated according to the following marginalization

\begin{equation}
P(\Omega_\Lambda)=\int  L(\Omega_\Lambda, \alpha,\beta, v_0, r_s) M(\Omega_\Lambda, \alpha,\beta, v_0, r_s)d\alpha d \beta d v_0 d r_s
\end{equation}

\begin{figure}
   \centering
\includegraphics[width=8cm]{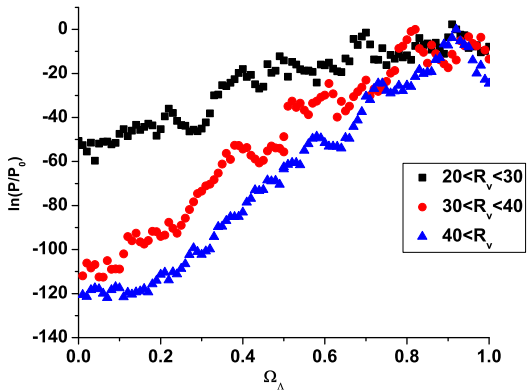}
\caption{ The probability ration vs. $\Omega_\Lambda$. We marginalize $\alpha$, $\beta$, $v_0$ and $r_s$ parameters, and assume the prior probability function is a constant. $P_0$ is the maximum probability. }
\label{like}
\end{figure}

 Here $M$ is a prior probability function. It depends on several parameters, but in general $M$ is assumed to be a constant. Figure \ref{like} shows the likelihood ratio under this assumption. Again the small $\Omega_\Lambda$s are ruled out immediately. But we must emphasize that the minimum $\chi ^2$s appearing in $30\mbox{Mpc}<R_v<40\mbox{Mpc}$ and $40\mbox{Mpc}<R_v$ cases are not reasonable and therefore $M$ cannot be taken as only a constant. A further study of the prior probability is needed.

\section{Conclusion}

In this paper, we studied the properties (orientation and density profiles) of the cosmological voids using the SDSS10 data.
We first confirmed that the voids' orientation is not purely random as one might naively expect. This effect has been first noticed by Foster etc\citep{Foster2009}. Later, Sutter etc. have found a constant empirical stretching factor using numerical simulations, and applied the stretching factor to SDSS DR10 data\citep{Sutter:2014oca,Lavaux}. They found that the voids' orientation can be used as a dark energy probe. Along these lines, a constant stretching factor will make the study of Alcick-Paczynski effect relatively straightforward. However, as illustrated in figure \ref{lambda-change}, we showed that the stretching factor highly depends on the coordinate which is used in the void finder\citep{Nadathur:2013bba}. This is not surprising, because different void finders give different void galaxies. Because of this discrepancy, it is very difficult to study the Alcick-Paczynski effect directly through the void orientation. To reduce this discrepancy, we included the galaxies outside voids into our analysis. Since the stretching effect is caused by the peculiar velocity, it is more realistic to apply a model which can quantify the peculiar velocity effect. It has been pointed out by Hamaus etal. that voids' density profiles are universal both inside and outside voids and their average velocities satisfy linear models very well\citep{Hamaus2014}. Therefore we apply Hamaus's density profile to the voids density profile in SDSS 10DR.

Although it is ultimately better to stack all the available data at once, figure \ref{density-profile} shows that the density profile also depends on the void's effective radius. This reduces the amount of voids that can be stacked together at once. In our study we separates the voids into three different effective radius intervals and fit the void density profiles with the empirical density distribution form. The result is consistent with the $\chi^2$ test and it implies that the empirical density profile form we obtained is reasonable. However, this is not enough to apply the result to calculate the confidence level directly, since the prior functions must be known. Unfortunately, in this case assuming a constant prior function is not suitable. It is unclear if this is caused by the limitation if the voids' number or by the voids' density profiles. This indicate that further analysis is warranted in order to better understand the voids' statistics and associated phenomena.

\section*{Acknowledgments}
D.C Dai was supported by the National Science Foundation of China (Grant No. 11433001), National Basic Research Program of China
(973 Program 2015CB857001), No.14ZR1423200 from the Office of Science and Technology in Shanghai Municipal Government and the key laboratory grant from the Office of Science and Technology in Shanghai Municipal Government (No. 11DZ2260700).

\label{lastpage}


\begin{thebibliography}{99}




\bibitem[Alcock et al.(1979)]{Alcock} Alcock C. and  Paczynski B.  1979, nat, 281, 358

\bibitem[Biswas et al.(2010)]{Biswas} Biswas, R., Alizadeh,
E., \& Wandelt, B.~D. 2010, prd, 82, 023002

\bibitem[Bos et al.(2012)]{Bos2012} Bos, E.~G.~P., van de
Weygaert, R., Dolag, K., \& Pettorino, V. 2012, mnras, 426, 440

\bibitem[Cai et al.(2015)]{2015MNRAS.451.1036C} Cai, Y.-C., Padilla, N., 
\& Li, B.\ 2015, mnras, 451, 1036 

\bibitem[Clampitt et al.(2013)]{Clampitt2013} Clampitt, J., Cai,
Y.-C., \& Li, B.\ 2013, mnras, 431, 749

\bibitem[Colberg et al.(2005)]{Colberg:2004nd} Colberg, J.~M., Sheth,
R.~K., Diaferio, A., Gao, L., \& Yoshida, N.\ 2005, mnras, 360, 216

\bibitem[Foster
\& Nelson(2009)]{Foster2009} Foster, C., \& Nelson, L.~A.\ 2009, apj, 699, 1252

\bibitem[Hamaus et al.(2014)]{Hamaus2014} Hamaus, N., Sutter,
P.~M., \& Wandelt, B.~D.\ 2014, Physical Review Letters, 112, 251302

\bibitem[Jennings et al.(2013)]{Jennings2013} Jennings, E., Li, Y.,
\& Hu, W.\ 2013, mnras, 434, 2167

\bibitem[Komatsu et al.(2011)]{komatsu} Komatsu, E., Smith,
K.~M., Dunkley, J., et al.\ 2011, apjs, 192, 18

\bibitem[Lavaux
\& Wandelt(2012)]{Lavaux} Lavaux, G., \& Wandelt, B.~D.\ 2012, apj, 754, 109

\bibitem[Lee \& Park(2006)]{Lee:2006gj} Lee, J., \& Park, D.\ 2006, apj, 652, 1

\bibitem[Lee
\& Pen(2000)]{Lee:1999ii} Lee, J., \& Pen, U.-L.\ 2000, apjl, 532, L5


\bibitem[Li
\& Zhao(2009)]{Li2009} Li, B., \& Zhao, H.\ 2009, prd, 80, 044027

\bibitem[Li et al.(2012)]{Li2012} Li, B., Zhao, G.-B.,
\& Koyama, K.\ 2012, mnras, 421, 3481

 \bibitem[Nadathur
\& Hotchkiss(2014)]{Nadathur:2013bba} Nadathur, S., \& Hotchkiss, S.\ 2014, mnras, 440, 1248

\bibitem[Nadathur et al.(2014)]{Nadathur:2014qja} Nadathur, S.,
Hotchkiss, S., Diego, J.~M., et al.\ 2014, arXiv:1407.1295

\bibitem[Padilla et al.(2005)]{Padilla:2005ea} Padilla, N.~D.,
Ceccarelli, L., \& Lambas, D.~G.\ 2005, mnras, 363, 977

\bibitem[Paz et al.(2013)]{Paz:2013sza} Paz, D., Lares, M.,
Ceccarelli, L., Padilla, N., \& Lambas, D.~G.\ 2013, mnras, 436, 3480

\bibitem[Park
\& Lee(2007)]{Park:2007qc} Park, D., \& Lee, J.\ 2007, apj, 665, 96

\bibitem[Peebles(1976)]{Peebles} Peebles, P.~J.~E.\ 1976, apj,
205, 318

\bibitem[Planck Collaboration et
al.(2014)]{Planck2013} Planck Collaboration, Ade, P.~A.~R., Aghanim, N., et al.\ 2014, aap, 571, AA16


\bibitem[Platen et al.(2008)]{Platen:2007ng} Platen, E., van de
Weygaert, R., \& Jones, B.~J.~T.\ 2008, mnras, 387, 128

 \bibitem[Porciani et al.(2002)]{Porciani:2001er}  Porciani, C., Dekel,
A., \& Hoffman, Y.\ 2002, mnras, 332, 339

\bibitem[Ricciardelli et al.(2013)]{Ricciardelli:2013kxa} Ricciardelli, E.,
Quilis, V., \& Planelles, S.\ 2013, mnras, 434, 1192

\bibitem[Ricciardelli et al.(2014)]{Ricciardelli:2014vga} Ricciardelli, E.,
Quilis, V., \& Varela, J.\ 2014, mnras, 440, 601

\bibitem[Ryden
\& Melott(1996)]{Ryden:1995tc} Ryden, B.~S., \& Melott, A.~L.\ 1996, apj, 470, 160

\bibitem[Sachs
\& Wolfe(1967)]{Sachs1967ApJ} Sachs, R.~K., \& Wolfe, A.~M.\ 1967, apj, 147, 73


\bibitem[Springel et al.(2005)]{2005Natur.435..629S} Springel, V., White,
S.~D.~M., Jenkins, A., et al.\ 2005, nat, 435, 629

\bibitem[Sutter et al.(2012a)]{Sutter2012} Sutter, P.~M., Lavaux,
G., Wandelt, B.~D., \& Weinberg, D.~H.\ 2012a, apj, 761, 44

\bibitem[Sutter et al.(2012b)]{Sutter2012a} Sutter, P.~M., Lavaux,
G., Wandelt, B.~D., \& Weinberg, D.~H.\ 2012b, apj, 761, 187



\bibitem[Sutter et al.(2014a)]{Sutter:2013pna} Sutter, P.~M., Lavaux,
G., Wandelt, B.~D., et al.\ 2014a, mnras, 442, 3127

\bibitem[Sutter et al.(2014b)]{Sutter:2013ssy} Sutter, P.~M., Lavaux,
G., Hamaus, N., et al.\ 2014b, mnras, 442, 462

\bibitem[Sutter et al.(2014c)]{Sutter:2014oca} Sutter, P.~M., Pisani,
A., Wandelt, B.~D., \& Weinberg, D.~H.\ 2014c, mnras, 443, 2983

\bibitem[Sutter et al.(2015)]{Sutter:2014haa} Sutter, P.~M., Lavaux, G., Hamaus, N., et al.\ 2015, Astronomy and Computing, 9, 1

\bibitem[Thompson \& Gregory(2011)]{Thompson2011} Thompson, L.~A., \& Gregory, S.~A.\ 2011, arXiv:1109.1268



\end{thebibliography}
\end{document}